\documentstyle[12pt,aaspp4]{article}  

\begin{document}

\title{\bf The Nearest Neighbor Alignment of Cluster X-ray Isophotes}
\author{Scott W. Chambers\altaffilmark{1,3},Adrian L. Melott\altaffilmark{1,4}
and Christopher J. Miller\altaffilmark{2,5}}
\altaffiltext{1}{Dept. of Physics \& Astronomy, Univ. of Kansas, Lawrence, KS
66045}
\altaffiltext{2}{Dept. of Physics \& Astronomy, Carnegie Mellon Univ.,
Pittsburgh, PA 15213}
\altaffiltext{3}{willc@kusmos.phsx.ukans.edu}
\altaffiltext{4}{melott@kusmos.phsx.ukans.edu}
\altaffiltext{5}{chrism@fire.phys.cmu.edu}

\begin {abstract}
We examine the orientations of rich galaxy cluster X-ray isophotes with
respect to their rich nearest neighbors using existing samples
of Abell cluster position angles
measured from {\it Einstein} and {\it ROSAT}
observations.  We study a merged subset of these samples using updated and
improved
positions and redshifts for Abell/ACO clusters.
We find high confidence for alignment, which
increases as nearest neighbor distance is restricted.
We conclude that there is a strong alignment signal in all this data,
consistent with gravitational instability acting on Gaussian perturbations.
\end {abstract}

\keywords{clusters of galaxies - Xrays: general --- large-scale structure of
universe}

\section{Introduction}
A conventional picture suggests that galaxy clusters form by hierarchical
clustering
whereby  material (gas, galaxies etc.) flows into denser regions along
interconnecting
large-scale filamentary structures (Shandarin \& Klypin 1984).
As a result of this infall,
clusters, in both optical and X-ray, appear aligned with their neighbors,
especially
when members of the same supercluster.
Observationally, this picture has been supported by previous alignment studies
(see
below), and dynamical evidence of relic drainage along supercluster filaments
(Novikov {\it et al.} 1999).

Most numerical simulations of structure formation by gravitational instability
acting on Gaussian initial perturbations
predict cluster alignment on some scale (e.g. Splinter et al. 1997;
Onuora \& Thomas 2000).
These simulations show that cluster alignments are not crucial for
discriminating between cosmological
models, but they support the gravitational instability hypothesis of structure
formation.

\subsection{Optical Alignments}
The projected shapes of galaxy clusters
on the sky are often elongated images (Carter \& Metcalfe 1980)
which can be approximated as ellipses.
The major axes of these define projected position angles,
measured counter-clockwise from north.
Binggeli (1982) found that the major axes of galaxy clusters tend to point
toward
their nearest neighbor.  Since then there have been many optical studies of
cluster alignments.
Most of the literature finds alignment on some scale.
For example, Flin (1987) and Rhee \& Katgert (1987) found significant
nearest neighbor alignment for separations, d$_{n}$ $<$ 30 h$^{-1}$Mpc.
West (1989) found that clusters which reside within
the same supercluster are significantly aligned on scales
of 30 h$^{-1}$Mpc and possibly out to 60 h$^{-1}$Mpc.
Rhee, van Haarlem \& Katgert (1992) also detected alignment for
clusters within the same supercluster, but they did not find
significance for nearest neighbor alignment.  Plionis (1994)
found significant alignment for d$_{n}$ $<$ 30 h$^{-1}$Mpc,
with weaker signals for larger separations up to 60 h$^{-1}$Mpc.
On the other hand, Strubles \& Peebles (1985) did not detect a significant
alignment signal (however, see Argyles {\it et al.} 1986).

\subsection{X-ray Alignments}
Individual galaxies may not be the best tracers of the shape of a cluster.
Problems can arise from foreground/background contamination, as well as the
contribution
of a discrete noise term. However, it is
believed that the X-ray emitting gas within a cluster
traces its gravitational potential (Sarazin 1986).
X-ray morphology is, then, probably the best observable
for determining galaxy cluster shape and orientation.
Likewise, since optical alignment has been measured and is well supported,
X-ray isophotal alignment is crucial for understanding
nearest neighbor cluster alignment.

Unfortunately there have not been many X-ray alignment studies.
Ulmer, McMillan, \& Kowalski (1989--hereafter UMK) performed the first
alignment
study using {\it Einstein} data on 46 clusters and found no significant result.
However, Chambers, Melott, and Miller (2000--hereafter CMM) recently
re-examined the
UMK results using updated cluster positions and found a very strong signal.
Rhee, van Haarlem \& Katgert (RvHK 1992) used the X-ray images of clusters
in a study of supercluster member alignment.
They examined clusters within the same supercluster
and a significant signal for nearest neighbor alignment was not detected.
However, West (1989) convolved X-ray and optical data to find that clusters are
aligned
when they are members of the same supercluster.
Further, West, Jones \& Forman (1995) showed that the X-ray substructure within
clusters
tends to share the orientation with its local environment out to 10$h^{-1}$Mpc.
Of the above analyses, which all used {\it Einstein} X-ray imagery,
only UMK and RvHK did nearest neighbor alignment studies, and neither
found support for the hypothesis. However, CMM, West, and West, Jones and
Forman find support for the hypothesis that the shapes of galaxy clusters
are aligned with LSS environment (as traced by nearest neighbors).
The amount and quality of X-ray cluster data since the UMK and RvHK analyses,
along with the conflicting reports discussed above, prompts a new analysis
of the nearest neighbor cluster alignment with X-ray isophotes.

In this paper we present cluster alignment results
for three samples with previously determined projected position angles.
We merge and filter these to obtain a larger sample.
Our goal is to use a well-controlled sample to test whether
the X-ray emitting gas within rich clusters is aligned with its
nearest neighboring rich cluster.
One of our adopted samples was previously tested for
nearest neighbor alignment in cases where clusters were within the same
supercluster;
a significant signal was not detected (RvHK, see above).
The other samples we adopt have not
been tested for nearest neighbor alignment (Mohr, Evrard, Fabricant \& Geller
1996;
Kolokotronis, Basilakos, Plionis \& Georgantopoulos 2001).
In an earlier work (CMM),
we studied a subsample of cluster orientations previously examined
by UMK.  Although they did not find a signal
for nonuniform orientations,
we detected a significant signal for alignment by using the Wilcoxon rank-sum
test (WRS; Lehmann 1975).
As the WRS is more effective in testing for alignments,
we use it in the present study.
We use q$_{o}$ = 0 and h = 100H$_{o}$ km s$^{-1}$ Mpc$^{-1}$ throughout.

\section{Data and Analysis}
\subsection{The Cluster Samples}
We did a literature search for existing X-ray position angles.
Rhee, van Haarlem \& Katgert used 107 rich clusters of galaxies to
study alignments within superclusters.
Of these, 27 had X-ray position angles determined from
{\it Einstein} imagery (Rhee \& Latour 1991).
When assigning these angles, they focused on the entire
X-ray image within the largest circular region centered on the
peak of the X-ray emission.
Although significant alignment was detected for clusters within
the same supercluster, nearest neighbor alignment was not found.
We therefore adopt these 27 RvHK clusters as one of our samples.
Position angles found from a moments method and an inertia tensor method
are both available; we use the latter in this analysis.

Mohr, Evrard, Fabricant \& Geller (1996--MEFG hereafter) used a sample of 65
galaxy clusters observed by the {\it Einstein} IPC to constrain the
range of cluster X-ray morphologies.  The core of
the X-ray emission was emphasized in their analysis.
Emission-weighted cluster orientations were calculated from the Fourier
expansion
of the photon distribution.
In the present work we only use Abell clusters for which we have
accurate redshift information.
This leaves 54 MEFG clusters (see next section).

Kolokotronis, Basilakos, Plionis \& Georgantopoulos (2001--KBPG hereafter)
found significant correlations between the Abell-APM optical and {\it ROSAT}
X-ray shape parameters of 22 rich galaxy clusters.
They concluded that the dynamical state of clusters can be indicated
by either part of the spectrum.
Their X-ray analysis focuses on the inner region of the emission.
Position angles were calculated by an inertia tensor method.
We also use this sample of 22 KBPG clusters.

\subsection{Nearest Neighbor Sample}
We measure the nearest neighbor distances using rich ($R \ge 1$) clusters
from the Abell (1958) and Abell, Corwin, and Olowin (1989) catalogs.
To have a well controlled neighbor sample, we follow the strategy of
CMM.
As R=0 clusters are not part of Abell's statistical sample,
we only use R $\ge$ 1 clusters in our study.
Also, since more accurate redshifts are now available, we have updated
the redshifts of all clusters.
The redshifts and coordinates used
are mainly from the MX Northern Abell Cluster Survey (MX; Slinglend {\it et
al.} 1998;
Miller {\it et al.} 2001) and the ESO Nearby Abell Cluster Survey (Katgert et
al. 1996).
We use the richness values from the Abell catalog for the Abell/ACO overlap
region ($-27^{\circ} \le \delta \le -17^{\circ}$).
There are 336 rich (R $\ge$ 1) clusters within the
$|$b$|$ $\ge $ 30$^{o}$ and 0.012 $\le$ z $\le$ 0.10 boundary of this survey.
These clusters have on average 25 redshifts per cluster and 86\% have at
least two measured galaxy redshifts.

We search through
the 336 rich clusters for possible nearest neighbors.  We note that
there are not many clusters with d$_{n}$ $>$ 30h$^{-1}$Mpc.
A potential weakness of most previous
nearest neighbor studies is the neglect of the survey boundary location.
Therefore, we note the distance and direction to the neighbor,
and the distance to the boundary of the survey,
all with respect to the subject cluster.  Since a potential nearest neighbor
may be hidden
just outside the boundary, we disqualify any pair in which the distance to
the boundary was less than the distance to the nearest neighbor.

We would like to briefly discuss the advantages of our controlled samples.  By
using
$R \ge 1$ clusters, we are ensured that we have a statistically complete
subsample.
Miller et al. (1999) found that cluster mean redshifts based on only one galaxy
redshift
are often erroneous by $\sim $ 5 h$^{-1}$Mpc, which can throw off the distance
to the nearest neighbor and perhaps more importantly, the identification
and direction to the true nearest neighbor.
Thus, updating the cluster redshifts makes sure that we are using the correct
nearest neighbor (and nearest neighbor distance).  And finally, excluding cases
where the boundary
distance is closer than the neighbor also helps to prevent the identification
of an
incorrect nearest neighbor.

\subsection{Statistical Samples}
After we apply these stringent criteria to our data, we are left with three
subsamples.
Twenty seven of the RvHK clusters are rich and within the boundary of the
survey.  However,
in four cases, the boundary was closer than the nearest neighbor, leaving
23 RvHK clusters.   The sample of MEFG lost 25 clusters because of richness or
not being
within the survey volume.  Of the 29 remaining, six pairs did not obey our
nearest neighbor-boundary distance requirement. We therefore can use 23 MEFG
clusters.
Four of the original KBPG clusters did not obey the richness and boundary
criteria.
And five of the 18 remaining clusters had the boundary distance closer than the
neighbor.  That is, we have 13 KBPG clusters after our cuts.
In summary, our statistical subsamples for RvHK, MEFG and KBPG contain 23/27,
23/54 and 13/22
clusters, respectively -- see Table 2.

We merged these three subsamples to obtain a final controlled sample.
We start with the 59 objects in our subsamples.  Some clusters
appear in more than one subsample, so there are
actually 48 individual Abell clusters total.
That is, there are 11 Abell clusters with
two independently measured orientations.
Nine of these clusters are common to the MEFG and RvHK
samples and  the other two Abell clusters are in both MEFG and KBPG.
We have kept eight (1367, 1650, 1983, 2065, 2147, 2199, 3158 \& 3266)
out of these 11 clusters, discarding three
(1767, 2124 \& 2151) because of inconsistent orientations.
We define consistent orientations are $\Delta$$\theta$ $\le$  25$^o$.
The position angles of the eight
consistent clusters were tested and found to be highly correlated.
Thus, which of these position angle we choose to use should not effect our
results.
We alternately pick position angles by chronology in Abell number,
keeping three RvHK, three MEFG and the two KBPG.
This makes a total of 45 total clusters in our final merged sample.
We have verified that different choices for these eight clusters does not
significantly change our conclusions.

\subsection{Analysis}

We use published position
angles, $\theta$, measured counter-clockwise from north,
for our statistical samples.
The absolute value of the projected angular difference
between this angle and the direction to the nearest neighbor defines the
pointing angle, $\phi_{p}$,
0$^{o}$ $\le$ $|\phi_{p}|$ $\le$ 90$^{o}$.

We define alignment as the tendency for the pointing angles, $|\phi_{p}|$,
to be systematically smaller than they would be if distributed isotropically
over this interval.  The Wilcoxon rank-sum test (WRS; Lehmann 1975) tests
for this.  The null hypothesis of WRS is that the sample is not
systematically smaller or larger than the (uniform) parent population.
It has been common for the Kolmogorov-Smirnov (KS) test to be used in
alignment studies.  However, KS tests for {\it any} difference from the assumed
parent population, in this case a pointing angle distribution uniform
over its possible range.  For example, KS would show a signal if there
were an excess of angles at
45$^{o}$, but this is not alignment (systematically small angles).
By being sensitive to any deviation, KS has reduced sensitivity to the
particular deviation for which we are testing.  WRS, on the other hand,
places the angles in the test sample in rank order against the angles in
the parent population (comparison sample).  Examination of these ranks
allows a specific test for whether the sample tends to be significantly
larger or smaller.

We have run the WRS test twice on each data sample.  The first,
which we call "uniform", uses a control group
placed at equal intervals between 0$^{o}$ and 90$^{o}$ with a mean of
45$^{o}$.
In the second method, called "remix", the control is constructed by reassigning
cluster position angles randomly within the sample, and
new pointing angles constructed using these.  The purpose of this
second method is to detect any possible systematic preference for a
given angle as a source of alignment.  Of course, larger fluctuations
(by $\sqrt[]{2}$)
are expected using this method.
We show the average of five remix computations.
The difference between the remix and uniform methods is barely more than
1$\sigma$ for this size remix population.

Table 1 shows the results of WRS on the merged and
re-mixed samples.  We have included D$_{max}$, the number of clusters,
N$_{cl}$ within D$_{max}$, and the probability that the alignment could
have arisen as a random sample of the control parent population.

We tested our merged sample for nearest
neighbor alignment at various maximum nearest neighbor
distances, D$_{max}$.  When D$_{max}$ $\ge 30$ h$^{-1}$Mpc signals
for alignment are low.  However, restricting D$_{max}$ $\le 20$ h$^{-1}$Mpc,
we find a significant signal for alignment with a 1.26\% probability for
no alignment.  The maximum signal for
alignment occurs at D$_{max}$ $\le 10$ h$^{-1}$Mpc where the probability
for no alignment is only 0.18\%.

\section{Conclusion and Discussion}

We tested three samples of previously determined
X-ray position angles, 45 clusters total, for nearest neighbor alignment.
We found, using the Wilcoxon rank-sum (WRS) test,
that the X-ray isophotes in our samples
are aligned with their nearest neighboring
rich Abell cluster, consistent with the gravitational instability
hypothesis (Shandarin \& Klypin 1984).
Alignment significantly increases
as the maximum neighbor distance is decreased.
In fact, we noticed a transition zone, before which isophotal alignment
becomes significant.
Our data shows that the orientation of the X-ray emitting gas within
rich clusters is correlated with nearest rich neighbor positions when d$_{n}$
$\le$ 20 h$^{-1}$Mpc;
this distance roughly marks a transition zone from orientations being random to
aligned.
This transition distance is the same found by Novikov, {\it et al.} (1999)
for alignment between cluster winds and the neighboring cluster population.
It likely represents the wavelength undergoing collapse at this time in cosmic
history, the so-called ``pancaking scale''
(Melott \& Shandarin 1993; Shandarin {\it et al.} 1995).

Previously, Rhee, van Haarlem and Katgert had not detected nearest neighbor
alignment in
their sample which we adopted.  If alignment were not present in our subset of
23 RvHK clusters,
then this would work against our signal for alignment - especially since this
is such a
large fraction of our merged sample.  However, we do not see this.  Instead, we
find a
fairly significant signal in our data.  We therefore speculate that updated
positions
and redshifts, as well as using a more controlled sample,
(particularly our boundary distance condition) allowed us to get a signal
for alignment from the RvHK data.
In a previous work (CMM 2000) we also showed the X-ray data of Ulmer, McMillan
and Kowalski (UMK 1989)
exhibits alignment, contrary to their conclusions.  Using a consistent data
analysis and a clear
statistical methodology, we have removed two crucial
discrepancies from X-ray alignment studies; now, all samples tested (to our
knowledge) have been shown to
exhibit (or contribute to) significant X-ray alignment on similar scales.

\section{Acknowledgments}

ALM and SWC are grateful for financial support under NSF grant
AST-0070702 and computing resources from the National Center for
Supercomputing Applications.

\begin{figure*}
\plotone{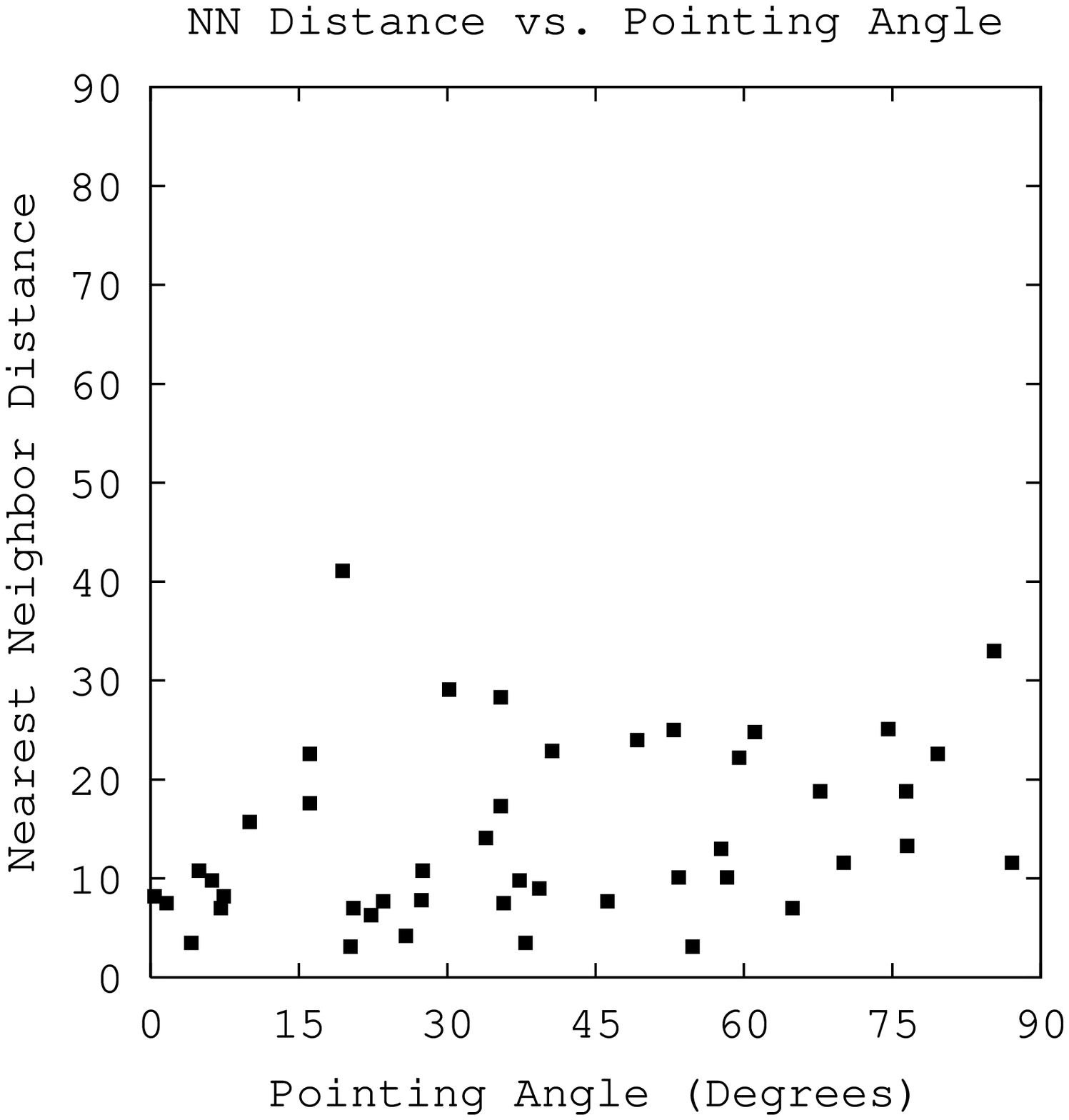}
\caption[]{\footnotesize
The distribution of pointing angle against nearest neighbor distance is shown
for our 45 clusters in our merged sample.  Pointing angles are generally $<$
 45$^o$ for separations $<$ 20 h$^{-1}Mpc$. }
\end{figure*}

{\footnotesize
\begin{deluxetable}{cccc}
\tablewidth{0pt}
\tablecaption{WRS Results for the Statistical Samples}
\tablehead{
\colhead{}  & \colhead{} & \colhead{Uniform}  & \colhead{Remix}\\
\colhead{D$_{max}$} &  \colhead{N$_{cl}$} & \colhead{Prob.} &
\colhead{Prob.}\nl
\colhead{($h^{-1}$Mpc)} & \colhead{} & \colhead{\%} & \colhead{\%} }
\startdata
10  &  18	&  0.18 	&0.99 \nl
20  &  32	&  1.26 	&1.66 \nl
30  &  43	&  7.36 	&6.55 \nl
40  &  44	&  10.75	&9.51 \nl
\enddata
\end{deluxetable}}

{\footnotesize
\begin{deluxetable}{cccccccc}
\tablewidth{0pt}
\tablecaption{Cluster Information}
\tablehead{
\colhead{Source} &  \colhead{Abell} & \colhead{z} & \colhead{$\theta$} &
\colhead{NN Abell} &
\colhead{z} &  \colhead{$d_{n}$} & \colhead{$\phi$} \nl
\colhead{} & \colhead{} & \colhead{} & \colhead{} & \colhead{} & \colhead{} &
\colhead{} & \colhead{} }
\startdata
     MFEG &     85 &   0.0556 &   169.0 &     87 &   0.0543 &      4.2 &
25.8\nl
     MFEG &    119 &   0.0444 &     2.0 &    168 &   0.0448 &     11.6 &
70.1\nl
     MFEG &    168 &   0.0448 &   165.0 &    119 &   0.0444 &     11.6 &
87.1\nl
     MFEG &    399 &   0.0722 &    16.0 &    401 &   0.0748 &      8.2 &
7.4\nl
     MFEG &    401 &   0.0748 &    23.0 &    399 &   0.0722 &      8.2 &
0.4\nl
     KBPG &    500 &   0.0666 &   162.3 &    514 &   0.0720 &     18.8 &
67.7\nl
     KBPG &    514 &   0.0720 &   126.8 &    500 &   0.0666 &     18.8 &
76.4\nl
     RvHK &   1185 &   0.0326 &    99.0 &   1228 &   0.0352 &     13.3 &
76.5\nl
     RvHK &   1291 &   0.0530 &    62.0 &   1377 &   0.0514 &      7.5 &
35.7\nl
     MFEG &   1367 &   0.0214 &   145.0 &   1656 &   0.0231 &     22.6 &
79.6\nl
     RvHK &   1367 &   0.0214 &   147.0 &   1656 &   0.0231 &     22.6 &
81.6\nl
     RvHK &   1377 &   0.0514 &    96.0 &   1291 &   0.0530 &      7.5 &
1.6\nl
     MFEG &   1644 &   0.0474 &    69.0 &   3555 &   0.0494 &     33.0 &
85.3\nl
     MFEG &   1650 &   0.0845 &   174.0 &   1663 &   0.0827 &      7.7 &
46.2\nl
     RvHK &   1650 &   0.0845 &   174.0 &   1663 &   0.0827 &      7.7 &
46.2\nl
     MFEG &   1656 &   0.0231 &    80.0 &   1367 &   0.0214 &     22.6 &
16.1\nl
     MFEG &   1767 &   0.0701 &    24.0 &   1904 &   0.0708 &     46.4 &
52.7\nl
     RvHK &   1767 &   0.0701 &   125.0 &   1904 &   0.0708 &     46.4 &
26.3\nl
     MFEG &   1775 &   0.0696 &   123.0 &   1795 &   0.0622 &     22.9 &
40.6\nl
     RvHK &   1795 &   0.0622 &    19.0 &   1831 &   0.0613 &      9.0 &
39.3\nl
     RvHK &   1809 &   0.0790 &   166.0 &   1780 &   0.0786 &     13.0 &
57.7\nl
     MFEG &   1983 &   0.0449 &   111.0 &   2040 &   0.0457 &     25.0 &
42.0\nl
     RvHK &   1983 &   0.0449 &     4.0 &   2040 &   0.0457 &     25.0 &
31.0\nl
     RvHK &   1991 &   0.0579 &   179.0 &   1913 &   0.0528 &     25.1 &
74.6\nl
     RvHK &   2029 &   0.0768 &    12.0 &   2028 &   0.0776 &      7.7 &
23.5\nl
     RvHK &   2040 &   0.0457 &    25.0 &   1983 &   0.0449 &     25.0 &
52.9\nl
     RvHK &   2063 &   0.0354 &   174.0 &   2147 &   0.0351 &     22.2 &
59.5\nl
     MFEG &   2065 &   0.0723 &   153.0 &   2056 &   0.0746 &      7.8 &
27.4\nl
     RvHK &   2065 &   0.0723 &   152.0 &   2056 &   0.0746 &      7.8 &
26.4\nl
     RvHK &   2079 &   0.0656 &   170.0 &   2092 &   0.0670 &      9.8 &
37.3\nl
     RvHK &   2092 &   0.0670 &    33.0 &   2079 &   0.0656 &      9.8 &
6.2\nl
     MFEG &   2124 &   0.0654 &   177.0 &   2122 &   0.0663 &      2.7 &
51.7\nl
     RvHK &   2124 &   0.0654 &   131.0 &   2122 &   0.0663 &      2.7 &
5.7\nl
     RvHK &   2142 &   0.0899 &   147.0 &   2178 &   0.0928 &     29.1 &
30.2\nl
     MFEG &   2147 &   0.0351 &   178.0 &   2151 &   0.0368 &      6.3 &
23.3\nl
     RvHK &   2147 &   0.0351 &   179.0 &   2151 &   0.0368 &      6.3 &
22.3\nl
     MFEG &   2151 &   0.0368 &   124.0 &   2152 &   0.0374 &      3.1 &
54.8\nl
     RvHK &   2151 &   0.0368 &   102.0 &   2152 &   0.0374 &      3.1 &
76.8\nl
     RvHK &   2152 &   0.0374 &    19.0 &   2151 &   0.0368 &      3.1 &
20.2\nl
     RvHK &   2197 &   0.0308 &   173.0 &   2199 &   0.0299 &      3.5 &
4.1\nl
     MFEG &   2199 &   0.0299 &    35.0 &   2197 &   0.0308 &      3.5 &
37.9\nl
     RvHK &   2199 &   0.0299 &    35.0 &   2197 &   0.0308 &      3.5 &
37.9\nl
     MFEG &   2410 &   0.0809 &    72.0 &   2377 &   0.0825 &     17.6 &
16.1\nl
     MFEG &   2420 &   0.0848 &    61.0 &   2428 &   0.0846 &     14.1 &
33.9\nl
     MFEG &   2657 &   0.0404 &    90.0 &   2506 &   0.0289 &     41.1 &
19.4\nl
     MFEG &   2670 &   0.0759 &   130.0 &   2638 &   0.0825 &     24.8 &
61.1\nl
     KBPG &   2717 &   0.0490 &   143.1 &   4059 &   0.0460 &     10.1 &
53.4\nl
     KBPG &   2734 &   0.0620 &     7.8 &   2800 &   0.0640 &     24.0 &
49.2\nl
     KBPG &   3093 &   0.0830 &    70.3 &   3111 &   0.0780 &     17.3 &
35.4\nl
     KBPG &   3111 &   0.0780 &    28.3 &   3112 &   0.0750 &     10.8 &
27.5\nl
     KBPG &   3112 &   0.0750 &     5.7 &   3111 &   0.0780 &     10.8 &
4.9\nl
     KBPG &   3128 &   0.0600 &    53.8 &   3158 &   0.0602 &      7.0 &
64.9\nl
     KBPG &   3158 &   0.0602 &    98.8 &   3128 &   0.0600 &      7.0 &
20.5\nl
     MFEG &   3158 &   0.0602 &   108.0 &   3128 &   0.0600 &      7.0 &
11.3\nl
     KBPG &   3223 &   0.0600 &   144.4 &   3151 &   0.0662 &     28.3 &
35.4\nl
     KBPG &   3266 &   0.0605 &    45.8 &   3231 &   0.0570 &     15.7 &
10.0\nl
     MFEG &   3266 &   0.0605 &    68.0 &   3231 &   0.0570 &     15.7 &
32.2\nl
     KBPG &   3897 &   0.0698 &    93.3 &   2480 &   0.0710 &      7.0 &
7.1\nl
     KBPG &   4059 &   0.0460 &   148.0 &   2717 &   0.0490 &     10.1 &
58.3\nl
\enddata
\end{deluxetable}}

\end{document}